\documentclass[11pt,a4paper]{article}

\usepackage{mathpple}
\usepackage{graphicx}
\usepackage{url}
\usepackage{paralist}

\usepackage{geometry}
\usepackage{fancyhdr}
\begin{document}

\thispagestyle{plain}

\noindent \textbf{Preprint of:}\\
D. K. Gramotnev, S. J. Goodman and T. A. Nieminen\\
``Grazing-angle scattering of electromagnetic waves in
gratings with varying mean parameters: grating eigenmodes''\\
\textit{Journal of Modern Optics} \textbf{51}, 379--397 (2004)

\hrulefill

\begin{center}

\textbf{\LARGE
Grazing-angle scattering of electromagnetic waves in gratings with
varying mean parameters: grating eigenmodes}

{\Large
D. K. Gramotnev$^1$, S. J. Goodman$^1$ and T. A. Nieminen$^2$}

$^1$Applied Optics Program, School of Physical and Chemical Sciences,
Queensland University of Technology, GPO Box 2434, Brisbane,
Queensland 4001, Australia; e-mail: d.gramotnev@qut.edu.au

$^2$Centre for Biophotonics and Laser Science, Department of Physics,
The University of Queensland, St Lucia, Queensland 4072, Australia

\begin{minipage}{0.8\columnwidth}
\section*{Abstract}
A highly unusual pattern of strong multiple resonances for bulk
electromagnetic waves is predicted and analysed numerically in
thick periodic holographic gratings in a slab with the mean
permittivity that is larger than that of the surrounding media.
This pattern is shown to exist in the geometry of grazing-angle
scattering (GAS), that is when the scattered wave ($+1$ diffracted order)
in the slab propagates almost parallel to the slab (grating) boundaries.
The predicted resonances are demonstrated to be unrelated to resonant
generation of the conventional guided modes of the slab. Their physical
explanation is associated with resonant generation of a completely new
type of eigenmodes in a thick slab with a periodic grating. These new
slab eigenmodes are generically related to the grating; they do not
exist if the grating amplitude is zero. The field structure of these
eigenmodes and their dependence on structural and wave parameters is
analysed. The results are extended to the case of GAS of guided modes
in a slab with a periodic groove array of small corrugation amplitude
and small variations in the mean thickness of the slab at the array
boundaries.
\end{minipage}

\end{center}

\section{Introduction}

Grazing-angle scattering (GAS) is a strongly resonant type of scattering
in uniform strip-like slanted wide periodic gratings [1­4]. It is realized
when the scattered wave ($+1$ diffracted order for first-order GAS [1,2,4],
or $+2$ order for second-order GAS [3]) propagates almost parallel to the
front grating boundary, that is at a grazing angle to this boundary. Thus,
GAS is intermediate between extremely asymmetrical scattering (EAS) (which
occurs when the scattered wave propagates parallel to the grating boundaries
[5­8]) and conventional Bragg scattering in reflecting or transmitting
gratings (where the scattered wave propagates at a significant angle with
respect to the grating boundaries).

The main distinctive feature of GAS is a strong resonant increase in
amplitudes of the scattered wave ($+1$ order [1,2,4] or $+2$ order [3])
and incident wave (zero order) inside a wide slanted grating at a resonant
(grazing) angle of scattering. A grating is regarded to be wide if its
width $L$ is larger than a determined critical width $L_c$ [9,10].
In narrow gratings (i.e. for $L<L_c$) the GAS resonance does not exist [1­3].

One of the most unexpected features of the GAS resonance is that
it strongly increases in height and sharpness with increasing
grating amplitude and/or grating width [1­4]. A physical explanation
for this resonant behaviour with respect to angle of scattering has
been related to resonant generation of a special new type of grating
eigenmodes that are guided by the grating in the absence of any
conventional guiding effect in the structure [2,3].

Another interesting and practically important feature of EAS and GAS
is their unusual and strong sensitivity to small step-like variations
of mean structural parameters (e.g. mean dielectric permittivity) at
the grating boundaries [4,11]. This is because one of the main physical
mechanisms resulting in EAS and GAS is diffractional divergence of the
scattered wave (similar to divergence of a laser beam of finite
aperture) [1,6,7,9­11]. Since diffractional divergence can be
strongly affected by small variations in mean structural parameters
[12], EAS and GAS appear to be highly sensitive to such variations
[4,11]. The expected consequences of such a sensitivity for the
experimental observation and application of EAS and GAS have been
discussed in [4,11].

In particular, Gramotnev et al. [4] investigated GAS in a symmetric
structure with a holographic grating in a slab surrounded by identical
media with the permittivities that are smaller than the mean permittivity
in the grating region (slab). However, only very small variations in the
mean permittivity at the grating boundaries were considered, such that
the angle of total internal reflection for the wave in the slab is larger
than the angle of scattering corresponding to the GAS resonance [4].
As a result, the tolerance of GAS to small variations in the mean
permittivity has been determined [4].

In this paper, we shall
demonstrate the existence of new, highly unusual strong multiple
resonances in a slanted holographic grating in the geometry of GAS
in the presence of the conventional guiding effect, that is when
the mean permittivity in the grating region (slab) is larger than
outside it. These multiple resonances will be shown to exist only
in sufficiently wide gratings (with $L > L_c$) and for relatively
large variations in the mean permittivity at the grating boundaries,
such that the angle of total reflection for the wave in the slab is
smaller than the GAS resonant angle.

A new type of electromagnetic
mode guided by a slab with a periodic holographic grating will be
discovered theoretically. The field structure and propagation
parameters of these modes will be investigated. The multiple GAS
resonances in the considered structure will be explained by resonant
generation of the predicted new slab eigenmodes. The analysis for
bulk electromagnetic waves will be carried out by means of the
rigorous theory of GAS and EAS [2,8], based on the enhanced \textbf{T}-matrix
algorithm [13,14]. The extension of the theory to the case of GAS of optical
modes guided by a slab with a corrugated boundary will also be discussed.

\section{Structure and methods of analysis}

\begin{figure}[htb]
\centerline{\includegraphics[width=0.5\columnwidth]{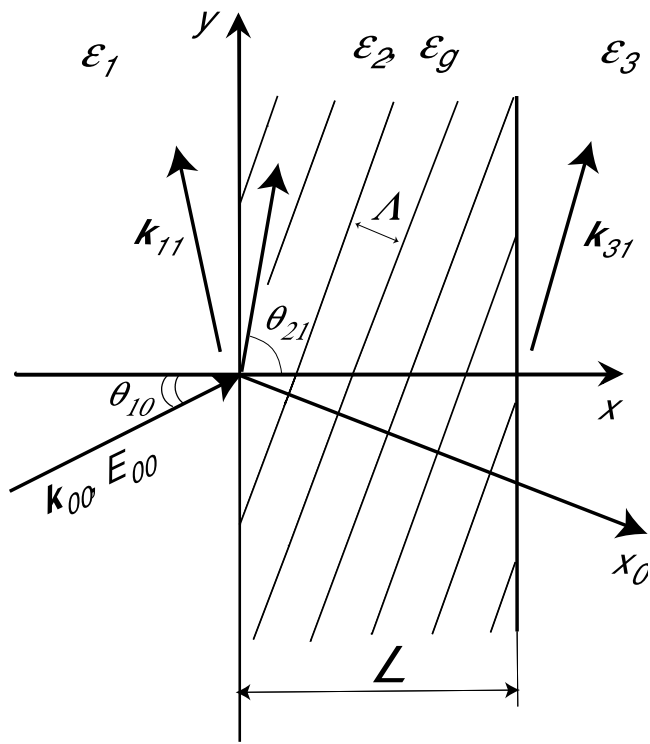}}
\caption{The scheme for GAS of bulk TE electromagnetic waves in
a slanted holographic grating of width $L$ and grating amplitude
$\epsilon_g$; $\epsilon_1$, $\epsilon_2$ and $\epsilon_3$ are the
mean permittivities in front, inside and behind the grating, respectively.
The angle of incidence is $\theta_{10}$, and the amplitude of the
incident wave in front of the grating is $E_{00}$. The Bragg condition
is satisfied precisely for the $+1$ diffracted order (scattered wave)
that propagates at an angle $\theta_{21}$ that is close to $\pi/2$
(i.e. at a grazing angle with respect to the grating boundaries),
and $\mathbf{k}_{11}$, $\mathbf{k}_{21}$ and $\mathbf{k}_{31}$
are the wave vectors of the scattered wave in front, in the middle
and behind the grating, respectively.}
\end{figure}

The structure that is under investigation in this paper is similar to
that analysed in [4]. Namely, we consider GAS of bulk transverse
electric (TE) electromagnetic
waves in a holographic grating in a slab of thickness $L$ between two
media of constant dielectric permittivities $\epsilon_1$ and
$\epsilon_3$ (figure 1):
\begin{equation}
\epsilon_s = \left \{ \begin{array}{l}
\epsilon_1,\\
\epsilon_2 + \epsilon_g \exp(\mathrm{i}Q_xx+\mathrm{i}Q_yy)
+ \epsilon_g^\ast \exp(-\mathrm{i}Q_xx-\mathrm{i}Q_yy) \\
\epsilon_3,
\end{array}, \textrm{for} \right \{ \begin{array}{l}
x<0,\\ 0<x<L,\\ x>L,
\end{array}
\end{equation}
where the coordinate system is shown in figure 1, $\epsilon_2$ is the
mean dielectric permittivity in the grating, $\epsilon_g$ is the
grating amplitude, $\mathbf{Q} = (Q_x, Q_y)$ is the reciprocal
lattice vector that is parallel to the $x_0$ axis,
$|\mathbf{Q}|=2\pi/\Lambda$,
and $\Lambda$ is the grating period. The grating is assumed to be
infinite along the $y$ and $z$ axes, and the dissipation is absent,
that is $\epsilon_{1,2,3}$ are real and positive. The step-like
variations of the mean dielectric permittivity at the front and
rear grating boundaries are given by
$\Delta\epsilon_1 = \epsilon_1 - \epsilon_2$ and
$\Delta\epsilon_2 = \epsilon_3 - \epsilon_2$.
In this paper we shall mainly assume that
$\Delta\epsilon_1 = \Delta\epsilon_2 < 0$, that is, the grating
region represents a planar waveguide capable of guiding
electromagnetic modes. A plane bulk TE electromagnetic wave
(with the amplitude $E_{00}$ and wave vector $\mathbf{k}_{00}$ in
medium 1 (see figure 1) is incident on to the grating at an angle
$\theta_{10}$ that is measured anticlockwise from the positive
direction of the $x$ axis to the vector $\mathbf{k}_{00}$ (figure 1).
The non-conical geometry of scattering is considered. The angle $\theta_{21}$
of scattering in the grating region is supposed to be close to $\pi/2$,
that is, the scattered wave (the $+1$ order) with
the wave vector $\mathbf{k}_{21}$ and the $x$-dependent amplitude
$S_{21}(x)$ propagates almost parallel to the grating boundaries
(the geometry of GAS) (figure 1).

As mentioned above, Gramotnev et al. [4] have investigated GAS in
the same structure, but with small variations in the mean permittivity
at the grating boundaries, such that the angle $\theta_{21\mathrm{int}}$
of total internal reflection for a wave in the slab was larger than the
angle $\theta_{21r}$ of scattering corresponding to the GAS resonance
in the same grating but with zero variations in the mean permittivity [4].
This condition restricted the analysis to the case of only very small
values of $\Delta\epsilon_1 = \Delta\epsilon_2 \approx 0.07
\epsilon_g \ll \epsilon_2$. This was important for understanding the
tolerances of GAS and its application possibilities for design of new
sensors and measurement techniques. At the same time, in the next
section we demonstrate that the case of larger variations in the
mean permittivity (when the angle of total reflection is larger
than the resonant angle for GAS) is even more interesting, since
it results in a discovery of a number of much stronger additional
resonances and a new type of eigenmodes of a slab with a slanted
holographic grating.

\section{Numerical results}

As mentioned above, Gamotnev et al. [4] determined the tolerance of
GAS to small symmetric variations in the mean dielectric permittivity
at the grating boundaries ($\Delta\epsilon_1 = \Delta\epsilon_2 <0$),
such that the angle $\theta_{21\mathrm{int}}$ of total internal
reflection for a wave in the slab was larger than or equal to the
GAS resonant angle $\theta_{21r}$ [1,2]. Figures 2 and 3 present the
dependences of amplitudes of the $+1$ order (scattered wave) on the
angle $\theta_{21}$ of scattering in the middle of the grating
(solid curves) and at the front and rear boundaries (dotted curves)
in the opposite case, that is for larger symmetric variations of the
mean permittivity ($\Delta\epsilon_1 = \Delta\epsilon_2 <0$),
such that $\theta_{21\mathrm{int}} < \theta_{21r}$. The specific
values of $\Delta\epsilon_1 = \Delta\epsilon_2 = 0.1508$ are chosen so
that $\theta_{21\mathrm{int}} = 80^\circ$, and we have a sufficiently
broad range of angles between $\theta_{21\mathrm{int}}$ and $\theta_{21r}$.
Note that for the considered structures the scattered wave amplitudes at
the front and rear grating boundaries are almost identical, and thus they
are represented by only one dotted curve in each of the subplots.
The grating (or slab) width is kept the same for all the subplots:
$L = 30 \mu$m (the other structural and wave parameters are presented
in the figure captions). The Bragg condition is assumed to be satisfied
precisely for all angles of scattering: that is, we investigate only
the effect of the angle of scattering on the pattern of GAS, excluding
the effect of frequency and/or angular detunings of the Bragg condition.

\begin{figure}[htb]
\centerline{\includegraphics[width=0.7\columnwidth]{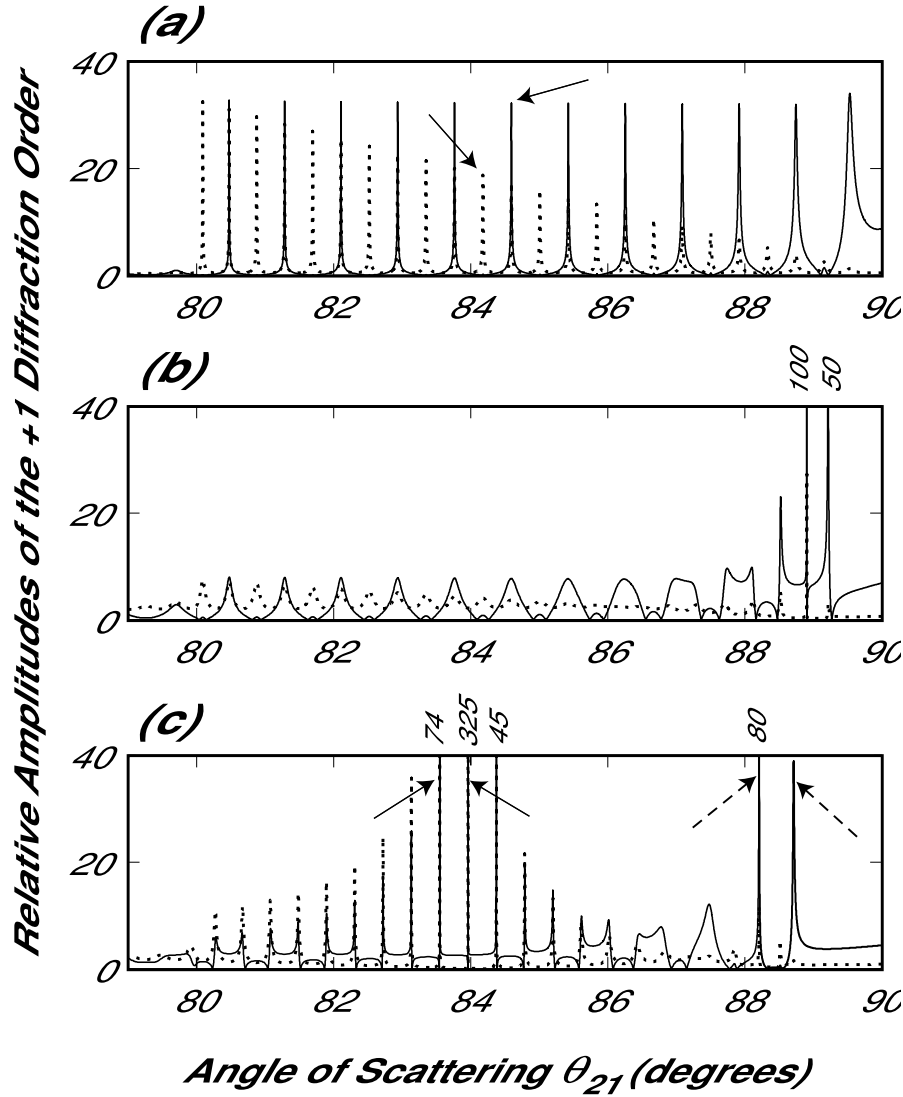}}
\caption{The dependences of relative amplitudes $|S_{21}/E_{00}|$
of the scattered wave (the $+1$ diffracted order) on scattering
angle $\theta_{21}$ in the middle of the grating (---) and at the
front and rear grating boundaries (...) for equal negative step-wise
variations in the mean dielectric permittivity at the grating
boundaries, $\Delta\epsilon_1 = \Delta\epsilon_2 = -0.1508$
($\Delta\epsilon_1 = \epsilon_1 - \epsilon_2$;
$\Delta\epsilon_2 = \epsilon_3 - \epsilon_2$), and different grating
amplitudes: (a) $\epsilon_g = 2\times 10^{-3}$ (the critical grating
width $L_c \approx 51 \mu$m); (b) $\epsilon_g = 8 \times 10^{-3}$
($L_c \approx 20 \mu$m), and (c) $\epsilon_g = 0.02$
($L_c \approx 11 \mu$m). The other structural parameters are:
$\epsilon_2 = 5$, $\theta_{10} = 45^\circ$, $L = 30 \mu$m, and
$\lambda(\mathrm{vacuum}) = 1 \mu$m. The numbers above the highest
resonances in (b) and (c) indicate their actual height.}
\end{figure}

\begin{figure}[htb]
\centerline{\includegraphics[width=0.7\columnwidth]{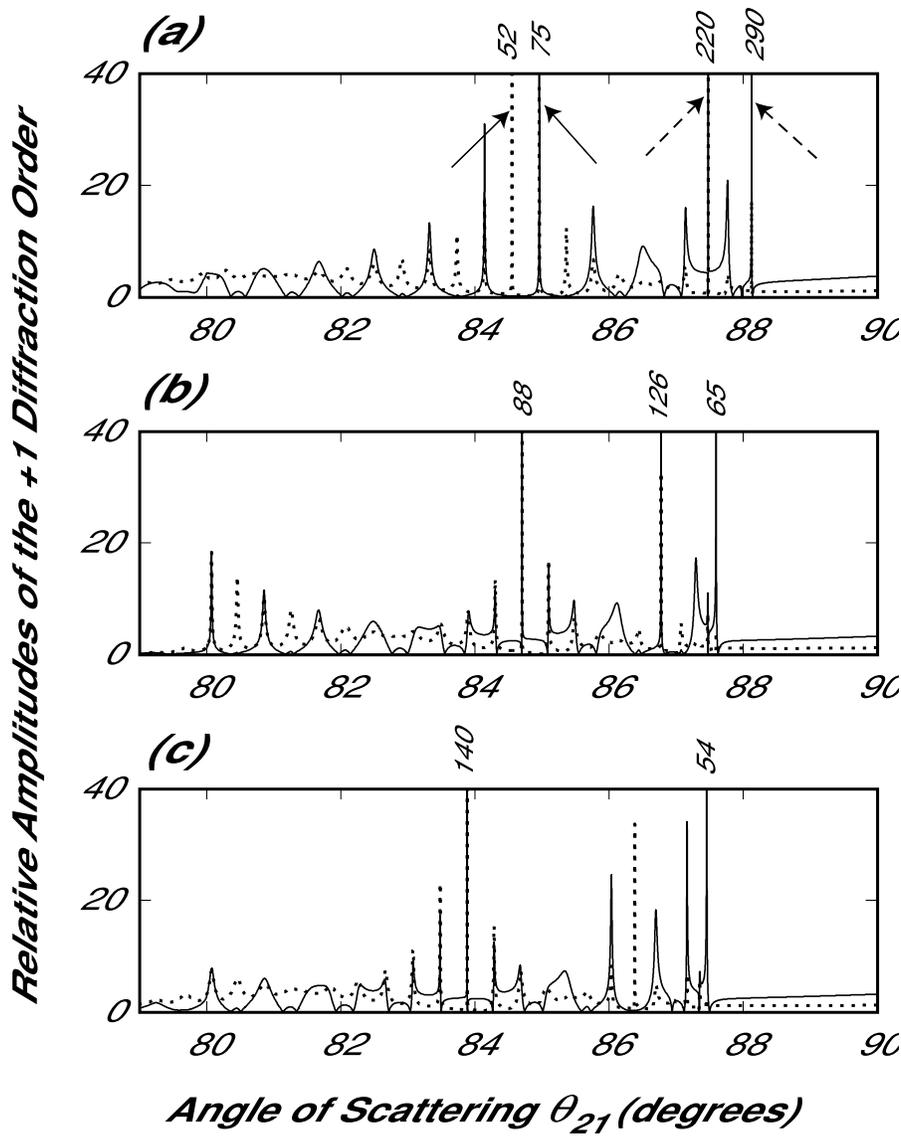}}
\caption{The same as for figure 2, but with (a) $\epsilon_g = 0.037$
($L_c \approx 9 \mu$m), (b) $\epsilon_g = 0.055$
($L_c \approx 7.2 \mu$m) and (c) $\epsilon_g = 0.06$
($L_c \approx 5.2 \mu$m).}
\end{figure}

Depending on the grating amplitude $\epsilon_g$, the considered grating
width $L = 30 \mu$m can be smaller or larger than the critical grating
width $L_c$. For example, for figure 2 (a), the grating amplitude
$\epsilon_g = 2 \times 10^{-3}$, which corresponds to
$L_c \approx 50 \mu$m [9]. This is noticeably larger than $L = 30 \mu$m.
Therefore, the pattern presented in figure 2 (a) is typical for narrow
gratings. It will be discussed in detail below. At this stage we shall
only mention that the presented resonances (figure 2(a)) are not related
to GAS, since GAS resonances do not exist in narrow gratings [1,2]
(for more detail see below).

Figures 2 (b) and (c) and 3 (a)--(c)
present the typical dependences of scattered wave amplitudes in the
middle of the grating (slab) and at its boundaries in wide gratings
of the same actual width ($L = 30 \mu$m), but with larger grating
amplitudes $\epsilon_g$ that make the critical grating width smaller,
that is $L > L_c$. The specific values of $L_c$ for each of the subplots
are presented in the figure captions.

It can be seen that increasing the grating amplitude so that the
grating width becomes just larger than $L_c$ (figure 2 (b)) results
in the appearance of extremely high and sharp resonances just below
$\theta_{21r}$ which is the GAS resonant angle in the same structure,
but with zero variations in the mean permittivity (for figure 2 (b),
$\theta_{21r} \approx 89.22^\circ$ [1,2]). The rightmost resonant
maximum in figure 2 (b) (i.e. in the structure with
$\Delta\epsilon_1 = \Delta\epsilon_2 <0$) occurs almost exactly at
$\theta_{21} \approx \theta_{21r}$. This statement is also correct
for all other subplots (figures 2 (c) and 3 (a)­(c)). Note that, as
mentioned in [1,2], $\theta_{21r}$ decreases with increasing grating
amplitude. Therefore, the rightmost maxima in figures 2 (b) and (d)
and 3 (a)­(c) shift to the left with increasing $\epsilon_g$.

Though the rightmost maxima in figures 2 (b) and (d) and 3 (a)­(c)
occur at $\theta_{21} \approx \theta_{21r}$, these resonances are
much stronger and sharper than those in the same
structures, but with zero variations in the mean permittivity.
For example, in figure 3 (a), the rightmost maximum goes up to about
$290E_{00}$, while in the same structure with
$\Delta\epsilon_1 = \Delta\epsilon_2 = 0$ it is only about $13E_{00}$.
Thus, symmetric negative variations in the mean permittivity at the
grating boundaries may result in an extremely strong increase in the
GAS resonance at $\theta_{21} = \theta_{21r}$. One of the physical
reasons for this effect is that, if $\theta_{21\mathrm{int}}<\theta_{21r}$,
then the scattered wave at $\theta_{21} \approx \theta_{21r}$ experiences
total internal reflection from the slab interfaces. This leads to a
substantial reduction in the energy losses from the grating, since the
$+1$ diffracted orders outside the grating become non-propagating
(evanescent) waves. As a result, the scattered wave amplitude inside
the grating may significantly increase.

In addition, if $L > L_c$, then on the left of the resonance at
$\theta_{21} \approx \theta_{21r}$ there appear a number of other
extremely strong and sharp (often, even sharper and stronger)
resonances at resonant angles $\theta_{21rs}$ within the angle range
$\theta_{21\mathrm{int}}<\theta_{21rs}<\theta_{21r}$ (figures 2 (b) and
(c) and 3 (a)­(c)). As the grating amplitude increases, so that the
grating width $L$ becomes just larger than $L_c$, a bunch of these
additional strong resonances appears just below the angle $\theta_{21r}$
(figure 2 (b)). If the grating amplitude is increased (so that $L$
becomes increasingly larger than $L_c$), this bunch of maxima `shifts'
from $\theta_{21r}$ towards $\theta_{21\mathrm{int}}$ (figure 2 (c)).
The word `shifts' is put in
quotes because the discussed resonant maxima do not actually move along
the horizontal axis with increasing $\epsilon_g$. Changing $\epsilon_g$
results in only insignificant variations in the values of the resonant
angles $\theta_{21rs}$. However, the height of these resonances can
change very strongly. For example, the sharp resonances within the
range between approximately $82^\circ$ and $85^\circ$ in figure 2 (c)
simply increase when the grating amplitude increases to $20\times 10^{-3}$.

When the first bunch of resonances `moves away' from the angle
$\theta_{21r}$ (for a sufficiently large $\epsilon_g$) another bunch
starts to appear near this angle (figure 2 (c)). This bunch also `shifts'
to the left towards $\theta_{21\mathrm{int}}$ with further increasing
$\epsilon_g$, and so on (figures 3 (a)­(c)). At larger grating amplitudes,
this process becomes less obvious, since the angular distance between the
bunches and the number of resonances in a bunch decreases
(figures 3 (a)­(c)). Note also that the number of resonances in a
bunch increases as the bunch `shifts' towards $\theta_{21r}$.

The resonances in the bunches are spaced (on the angular scale)
approximately periodically, especially for smaller grating amplitudes
(figures 2 (c) and 3 (a)). The period is approximately the same for all
the bunches, but resonances in each successive bunch occur approximately
between the resonances in the previous bunch (figures 2 (b) and (c) and
3 (a)­(c)). Increasing grating width results in a rapid increase in the
total number of the resonances, that is, the angular distance between
them is reduced. Increasing grating width also results in increasing
typical height and sharpness of the resonances. On the other hand,
changing values of $\Delta\epsilon_1 = \Delta\epsilon_2 <0$ has only a
limited effect on the values of $\theta_{21rs}$ and the typical angular
distance between the resonances. At the same time, varying the mean
permittivity in the grating can substantially change (increase or
decrease) the height and sharpness of the resonances (optimization
of the values of
$\Delta\epsilon = -\Delta\epsilon_1 = -\Delta\epsilon_2 >0$ will be
discussed below). The number of the GAS resonances is also increased
with increasing $\Delta\epsilon$, because of the increasing range of
angles between $\theta_{21\mathrm{int}}$ and $\theta_{21r}$.

If an asymmetric structure is considered (with
$\Delta\epsilon_1 \ne \Delta\epsilon_2$, $\Delta\epsilon_1 < 0$ and
$\Delta\epsilon_2< 0$), then some of the GAS resonances may strongly
increase, while others may be reduced, depending on particular values
of the variations of the mean permittivity. At the same time, the
angular position of the resonances depends on $\Delta\epsilon_1$ and
$\Delta\epsilon_2$ only very weakly. In this case, strong GAS resonances
can be observed within the range of angles
max($\theta_{21\mathrm{int}1},\theta_{21\mathrm{int}2}$)%
$<\theta_{21}<\theta_{21r}$, where $\theta_{21\mathrm{int}1}$ and
$\theta_{21\mathrm{int}2}$ are the angles of total internal reflection
from the front and rear boundaries of the grating (slab). Below this
range, strong GAS resonances cannot exist, because in this case the
scattered wave either in front (if $\Delta\epsilon_1 > \Delta\epsilon_2$)
or behind (if $\Delta\epsilon_1 < \Delta\epsilon_2$) the grating becomes
a propagating wave travelling away from the grating. This wave is
associated with substantial energy losses from the grating, resulting
in a deterioration of any strong resonance of scattering. This is also
the reason why the described pattern of strong multiple GAS resonances
(figures 2 (b) and (c) and 3 (a)­(c)) cannot exist if at least one of
the variations $\Delta\epsilon_1$ or $\Delta\epsilon_2$ is zero,
positive or sufficiently small negative that the angle of total
reflection from the corresponding slab interface is larger than
$\theta_{21r}$ [4].

\section{Physical reasons: conventional guided modes or not?}

It is obvious that the structure with
$\Delta\epsilon_1 = \Delta\epsilon_2 <0$ (i.e. the mean permittivity in
the slab is higher than that of the surrounding media) is capable of
supporting
the conventional guided modes. Therefore, a reasonable question
naturally arises from the above consideration: are all, or at least
some, of the predicted strong multiple GAS resonances in figures 2 (b)
and (c) and 3 (a)­(c) related to generation of the conventional guided
modes of the slab? To answer this question, let us consider physically
the conditions for effective generation of conventional guided modes
in a slab with a holographic grating. A conventional guided mode can
be represented by a bulk wave travelling in the slab and experiencing
total internal reflection from its boundaries [15]. An angle $\theta_{21sm}$,
at which this bulk wave should propagate in the slab to form a particular
slab mode, is determined by the dispersion relationship for the guided
mode and the fact that the wave vector of this guided mode must be equal
to the tangential (to the slab) component of the wave vector of the bulk
wave in the slab [15]. Thus, $\theta_{21sm}21sm = \arcsin(q/k_{21})$,
where $q$ is the magnitude of the wave vector of the considered
conventional slab mode. Determining q from the dispersion relationship
for the slab modes, we can easily see that each of the resonant angles
corresponding to the sharp resonances within the range from $80^\circ$
to $86^\circ$ in figure 2 (c) lies almost exactly in the middle between
two neighbouring values of $\theta_{21sm}$. This clearly suggests that
the above-mentioned GAS resonances cannot be associated with generation
of the conventional slab modes.

On the other hand, the same consideration
demonstrates that all the resonances in figure 2 (a) occur at the angles
corresponding to the conventional slab modes in the considered structure.
This suggests that, if the grating amplitude and/or grating width are
sufficiently small, we indeed have resonant generation of the conventional
slab modes in the structure (the maxima of the solid curve in figure 2 (a)
correspond to symmetric TE slab modes, while the maxima of the dotted
curve correspond to antisymmetric TE slab modes).

This is because a bulk
wave in a slab (reflecting from the slab interfaces) forms a guided mode
only if it experiences at least several reflections from the slab
interfaces, otherwise it is just a bulk wave travelling from one slab
boundary to another with no restrictions on the angle of propagation.
On the other hand, scattering in the grating is characterized by a
critical length of the grating (along the $y$ axis),
$l_c \approx \tau c \epsilon_2^{-1/2}$, within which the steady-state
regime of scattering is reached [7,16]. Here, $\tau$ is the relaxation
time for a particular resonance [7,16], and $c$ is the speed of light
in vacuum. Resonant generation of the conventional modes guided by a
slab with a holographic grating occurs (figure 2 (a)) when lc for the
corresponding resonance is noticeably larger than the distance
$l_\mathrm{ref} = L/\cos(\theta_{21})$ that the bulk wave in the slab
travels between two successive reflections. For example, for figure
2 (a), the analysis developed in [16] gives $l_c \approx 0.4$\,cm, while
$l_\mathrm{ref}$ ranges from about 0.02\,cm to 0.2\,cm for the angles
$\theta_{21}$ within the range between $80^\circ$ and $89^\circ$.

If the grating width is increased, then $l_\mathrm{ref}$ increases
proportionally. If $l_c<l_\mathrm{ref}$, we rather have generation
of a bulk wave in the grating region, and the successive reflections
do not effectively restrict angles of scattering so that they can vary
in a wide range, not necessarily corresponding to a conventional guided
mode. This may also happen when the grating amplitude is increased and
$L =$ constant. In this case, $l_\mathrm{ref}$ is constant, but the
efficiency of scattering is increased, resulting in a smaller critical
grating length $l_c$. This eventually leads to the same inequality
$l_c<l_\mathrm{ref}$. Therefore, the resonances due to generation of
the conventional guided slab modes must become broader with increasing
grating width $L$ and/or grating
amplitude $\epsilon_g$. This is what can clearly be seen from the
comparison of the curves in figure 2 (a) with the curves in figure 2 (b)
at angles $\theta_{21}<87^\circ$. (A very similar effect occurs when the
grating width is increased and  $\epsilon_g =$ constant).

Note that increasing angle of scattering results in increasing
$l_\mathrm{ref}$. Therefore, the maxima in figures 2 (a) and (b) due to
the conventional slab modes increase in width with increasing $\theta_{21}$.
Eventually, for angles of scattering $\theta_{21} \approx 87^\circ$ in
figure 2 (b), lc becomes approximately the same as $l_\mathrm{ref}$
and the whole pattern of scattering drastically changes; for angles
$\theta_{21} > 87^\circ$, the resonances due to the conventional guided
modes disappear, being replaced by strong GAS resonances at `wrong'
resonant angles of scattering (figure 2 (b)).

However, it is important to note that, for strong GAS resonances,
the relaxation time is large, and $l_c$ becomes again larger than
$l_\mathrm{ref}$. This is also quite clear from the fact that, in
the GAS resonances, the selectivity of scattering with respect to
angle $\theta_{21}$ is extremely high (figures 2 (b) and (c) and
3 (a)­(c)), which can only be achieved if the effect of the rear
slab boundary on scattering is significant (recall that we do not
consider the angular response of the grating, and the Bragg condition
is satisfied precisely for all scattering angles). That is, the bulk
scattered wave in the grating should be able to travel many times across
the grating, before the steady state is finally reached. This implies that
for the GAS resonances the condition $l_c>l_\mathrm{ref}$ must again
be satisfied.

As a result of this consideration, we can draw an important conclusion
that strong GAS resonances in a slab with a slanted holographic grating
have nothing to do with the resonant generation of the conventional slab
modes. All the predicted GAS resonances represent a new phenomenon in a
slab with a slanted grating.

This conclusion can be made even more
convincing if we investigate the angular dependences of the scattered wave
amplitude in the grating at angles $\theta_{21} > 90^\circ$. Indeed,
if we extend the curves in figure 2 (a) to angles $\theta_{21} > 90^\circ$,
they will be almost exactly symmetric with respect to
$\theta_{21} = 90^\circ$. This is well expected, since a conventional
guided slab mode can be represented by a bulk wave successively
reflecting from the slab interfaces. Therefore, scattering at two
angles $90 - \Delta\theta_{21}$ and $90 + \Delta\theta_{21}$ can equally
result in generation of the same mode guided by the slab.

On the contrary,
no GAS resonances can be observed at angles $\theta_{21} > 90^\circ$.
This is similar to what was predicted for gratings with zero variations
in the mean permittivity at the boundaries [1,2]. That is, unlike
scattering into conventional guided modes of a slab with a holographic
grating, the pattern of GAS is completely asymmetric with respect to
the scattering angle $\theta_{21} = 90^\circ$. This once again emphasizes
strong differences between the resonances of scattering due to generation
of conventional slab modes (figure 2 (a)) and GAS resonances (figures 2 (b)
and (c) and 3 (a)­(c)).

\section{New grating eigenmodes}

In the previous section, we have clearly demonstrated that the
predicted multiple-GAS resonances have nothing to do with generation
of the conventional guided modes in the slab (grating region). However,
this still does not give us a physical explanation for these new
resonances. On the other hand, it is well known that any strong
resonance is associated with some kind of eigenoscillations or
eigenmodes in the structure. Therefore, the described strong GAS resonances
must also be related to generation of some type of eigenmodes in the
considered structure with a grating in a slab. At the same time, as
demonstrated in the above section, these are not conventional guided
slab modes that are responsible for the predicted GAS resonances.
Therefore, a completely new type of eigenmode must be generated in
the slab by means of an incident wave. These are grating eigenmodes
that were first predicted in [2,3]. However, in [2,3], grating eigenmodes
were considered only in the absence of the conventional guiding effect
in the structure. Here, we demonstrate that the presence of the
conventional guiding effect, that is when the mean permittivity in the
grating region is larger than the permittivity outside the grating,
results in drastic changes in the pattern of GAS and therefore of the
structure of the grating eigenmodes.

Indeed, figures 2 and 3 clearly demonstrate that a large number of
very sharp and strong resonances can exist in a grating with
$\Delta\epsilon_1 = \Delta\epsilon_2 <0$. These resonances appear to
be much higher than those in the case when
$\Delta\epsilon_1 = \Delta\epsilon_2 =0$ [2] (figures 2 and 3). The
typical number of these sharp and strong resonances rapidly increases
with increasing grating width $L$ and/or mean permittivity in the grating
(see above), so does the number of different grating eigenmodes in the
structure.

It is also clear that the discussed grating eigenmodes in a slab are not
true eigenmodes, because they weakly leak into the regions outside the
grating (see also [2,3]). Indeed, these modes are resonantly generated
by an incident bulk wave, and coupling between guided and bulk waves can
occur only if the guided waves (grating eigenmodes) leak into the
surrounding media. At the same time, since the corresponding GAS
resonances are extremely high (up to hundreds (figures 2 and 3) or
even thousands (see below) of the amplitude of the incident wave at
the front grating boundary), the leakage of the grating eigenmodes
must be very weak (weak coupling between the incident wave and grating
eigenmodes). Therefore, these eigenmodes should be able to propagate long
distances along the grating (in the absence of the incident wave at the
front boundary), before their amplitudes are significantly decreased
owing to leakage. These typical distances can approximately be determined
by the relaxation time for the corresponding GAS resonances. The larger
the relaxation time, the stronger is the resonance and the longer is the
distance that the corresponding eigenmode can propagate without noticeable
decay (due to leakage) along the grating. The relaxation time can be
determined using the method based on the Fourier analysis of the incident
pulse [16]. As mentioned above, during the relaxation time $\tau$, the
scattered wave propagates the distance $l_c \approx \tau c \epsilon^{-1/2}$
along the grating, which should approximately be equal to the distance that
the corresponding eigenmode can propagate along the grating in the absence
of the incident wave. For example, the results of [16] suggest that a
resonance of about $100E_{00}$ high should usually correspond to an $l_c$
value of about several tens of centimetres.

It is important to understand
that the predicted new eigenmodes of a guiding slab with a grating are
generically associated with scattering. They do not exist if there is
no grating in the slab. This is one of their major differences from the
conventional slab modes. Grating eigenmodes exist only at sufficiently
large grating amplitudes and/or widths of the grating (see the previous
section and [2,3]). If the grating amplitude and/or grating width are
small (so that $L<L_c$), GAS resonances and the corresponding eigenmodes
do not exist and are replaced by the resonances caused by generation of
the conventional guided slab modes (figure 2 (a)).

\begin{figure}[t!pb]
\centerline{\includegraphics[width=0.7\columnwidth]{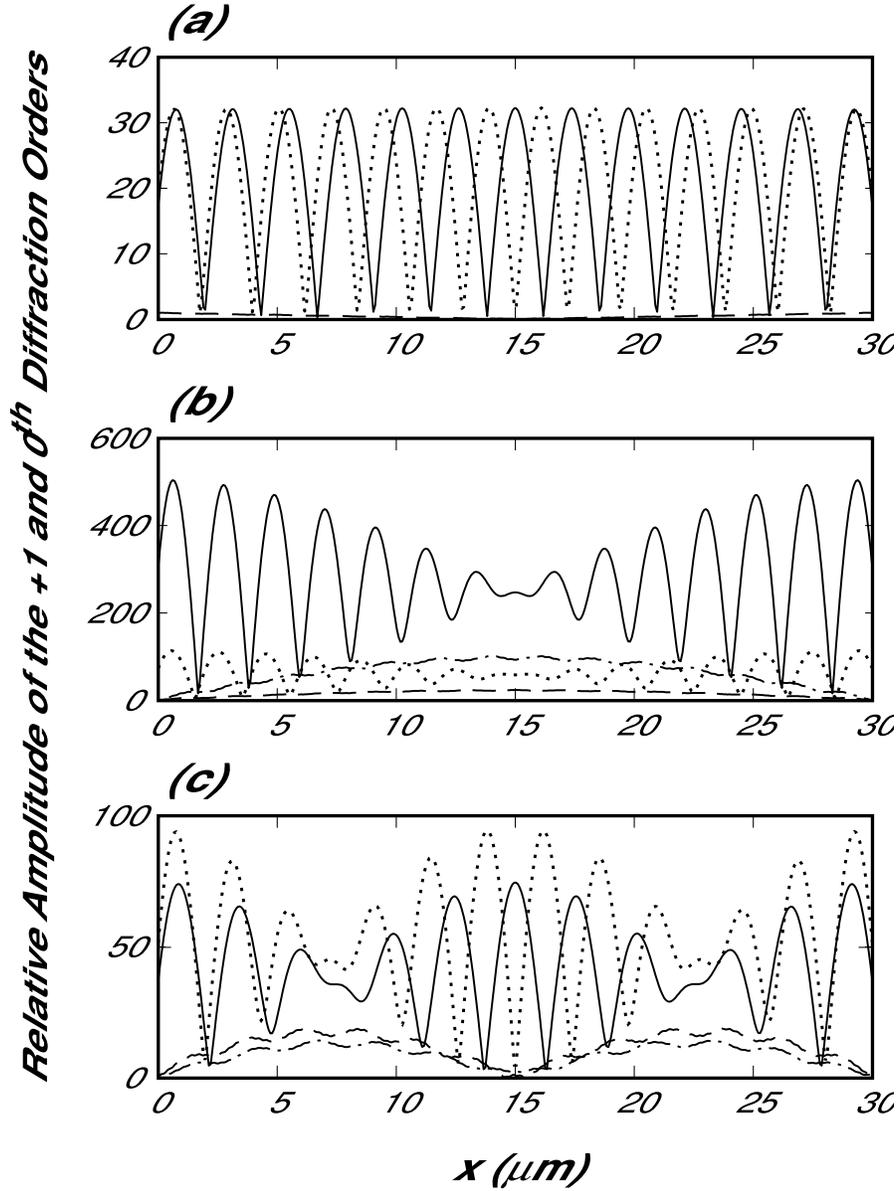}}
\caption{The typical dependences of the relative amplitudes
$|S_{21}(x)/E_{00}|$ (---,...) and $|S_{20}(x)/E_{00}|$ (-\,-\,-,-.-)
of the scattered and incident waves on distance $x$ from the front
grating boundary for the resonances indicated in figures 2 (a) and
(c) and 3 (a) by the solid arrows. (a) The $x$ dependences corresponding
to the resonances indicated by the arrows in figure 2 (a) (the zeroth
bunch of maxima); (...), $\theta_{21} = 84.1805^\circ$, (---),
$\theta_{21} = 84.596^\circ$. The dependences of the incident wave
amplitude are almost indistinguishable for both the resonances and
are given by just one broken curve. (b) The $x$ dependences corresponding
to the resonances indicated by the solid arrows in figure 2 (c)
(the first bunch); (...), (­\,­\,­): $\theta_{21} = 83.5497^\circ$,
(---)(­.­), $\theta_{21} = 83.963495^\circ$. (c) The dependences
corresponding to the resonances indicated by the solid arrows in
figure 3 (a) (the second bunch); (...), (­\,­\,­),
$\theta_{21} = 84.55485^\circ$, (---), (­.­), $\theta_{21} = 84.9622^\circ$.}
\end{figure}

The typical $x$ dependences of the relative amplitudes $|S_{21}(x)/E_{00}|$
and $|S_{20}(x)/E_{00}|$ of the scattered and incident waves, respectively,
inside the grating are presented in figure 4 for several strong resonances
that are indicated by the solid arrows in figures 2 (a), 2 (c) and 3 (a)
(other diffracted orders have negligible amplitudes at the considered
grating amplitudes). Since these resonances are very high, the field
structure in the grating at the corresponding resonant angles $\theta_{21rs}$
of scattering should approximately be the same as that of the corresponding
eigenmodes. Indeed, in this case, the modes are only weakly leaking from
the grating (slab), and the amplitude of the incident wave at the front
grating boundary is much smaller than the amplitudes of the scattered and
incident waves in the grating (figure 4). Therefore, the resultant sum of
the fields corresponding to the eigenmode and to the incident wave will
hardly be different from the field structure of the eigenmode alone.
Note that this is relevant to any sufficiently sharp resonance,
irrespective of whether this resonance is due to the conventional
guided slab modes (figure 2 (a)), or new grating eigenmodes (figures 2 (b)
and (c) and 3 (a)­(c)). Therefore, the dependences presented in
figures 4 (a)­(c) with a good approximation represent the field structure
of the corresponding eigenmodes (conventional and new).

Thus the dotted and
solid curves in figure 4 (a) approximately represent the field structure
in symmetric and antisymmetric conventional guided modes of the slab,
corresponding to the resonances indicated by the solid arrows in figure
2 (a). The broken curve represents the $x$ dependences of the incident
wave amplitude for both the considered resonances. However, the incident
wave in the grating is not a part of the conventional slab eigenmode.
Indeed, the incident wave amplitude in the grating can be made arbitrarily
small compared with the amplitude of the scattered wave by reducing the
grating amplitude. Therefore, conventional slab modes can obviously exist
in a slab with no grating.

The situation with the grating eigenmodes is completely different. Firstly,
grating eigenmodes are usually generated at different angles of scattering
(see section 3). Secondly, grating eigenmodes of a slab are intrinsically
related to scattering and cannot exist in the absence of the grating.
Thirdly, because they are generically related to scattering, the 0
diffracted order (incident wave) in the grating is an inseparable part
of the grating eigenmodes (also [2,3]). The amplitude of the 0 diffracted
order in the grating can never be made arbitrarily small compared with the
amplitude of the scattered wave ($+1$ order) without destroying the eigenmode
(and the corresponding GAS resonance). Thus the field structure in a grating
eigenmode is given not only by the scattered wave but rather by a
superposition of the fields of all diffracted orders involved (see
also [2,3]).

Figures 4 (b) and (c) present the typical dependences
of the scattered and incident wave amplitudes in the grating for the
several resonances indicated by the solid arrow in figures 2 (c) and
3 (a) (corresponding to generation of grating eigenmodes in the slab).
It can be seen that these dependences are significantly different from
those in figure 4 (a). In particular, unlike figure 4 (a), the amplitudes
of the incident wave in figures 4 (b) and (c) are much larger than $E_{00}$,
its amplitude at the front grating boundary. As mentioned above, there is
no way that we could make them negligible compared with the amplitude of
the scattered wave, without destroying the corresponding resonances and,
hence, the grating eigenmodes.

Comparison of figure 4 (a) with figures 4 (b) and (c) suggests that the
scattered wave amplitudes inside the grating experience strong oscillations
(as a function of $x$) for both conventional slab modes and grating
eigenmodes. However, the $x$ dependences corresponding to grating
eigenmodes display beat-like modulation of these oscillations with
one or more waists (see the solid and dotted curves in figures 4 (b)
and (c)). The number of these waists is determined by the bunch of
maxima (figures 2 (b) and (c) and 3 (a)­(c)) that the considered
resonance belongs to. For example, the $x$ dependences of the scattered
wave amplitude in figure 4 (b) show only one waist if the corresponding
resonances belong to the first bunch of maxima
(figure 2 (c)). The two waists are displayed by the solid and dotted
curves in figure 4 (c), because the corresponding resonances belong
to the second bunch of maxima (figure 3 (a)), etc.

The typical number
of oscillations of the $x$ dependences of the scattered wave amplitude
in the grating increases with decreasing resonant angle $\theta_{21rs}$
of scattering (i.e. with increasing grazing angle between $\mathbf{k}_{21}$
and the grating boundaries). For example, if we shift from the resonance in
figure 2 (c), indicated by the right solid arrow, to the neighbouring
resonance indicated by the left solid arrow, the number of oscillations
of the $x$ dependencies of the scattered wave amplitudes is increased by
one (compare the dotted and solid curves in figure 4 (b)). This is very
similar to what occurs for the conventional guided modes (figure 4 (a)).

\begin{figure}[!tb]
\centerline{\includegraphics[width=0.7\columnwidth]{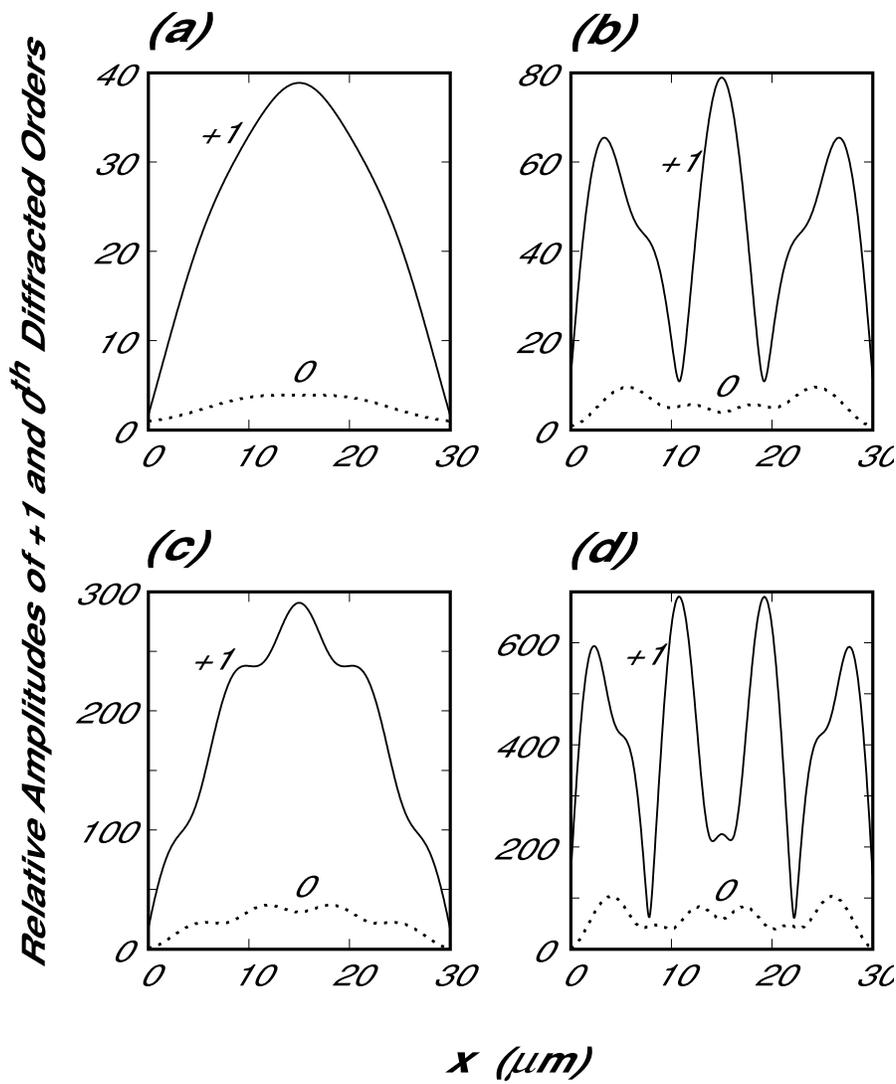}}
\caption{The typical dependences of the relative amplitudes of the
scattered waves (---) and incident waves (...) on distance from the
front grating boundary for the resonances near the angle $\theta_{21r}$,
indicated in figures 2 (c) and 3 (a) by the broken arrows. (a) The
resonance indicated by the right broken arrow in figure 2 (a), that
is $\theta_{21} \approx \theta_{21r} \approx 88.7143^\circ$; (b) the
resonance indicated by the left broken arrow in figure 2 (a), that
is $\theta_{21} \approx 88.2095^\circ$; (c) the resonance indicated
by the right broken arrow in figure 3 (a), that is
$\theta_{21} \approx \theta_{21r} \approx 88.12875^\circ$; (d)
the resonance indicated by the left broken arrow in figure 3 (a),
that is $\theta_{21} \approx 87.47955^\circ$.}
\end{figure}

Thus, if we consider the GAS resonance at $\theta_{21} \approx \theta_{21r}$
(the largest possible resonant angle), the number of oscillations of the $x$
dependence of the scattered wave amplitude in the grating is reduced to one.
This is demonstrated by figures 5 (a) and (c) that are plotted for the
rightmost resonances in figures 2 (c) and 3 (a), respectively. The number
of bumps on the solid curves in figures 5 (a) and (c) (two for figure 5 (a)
and four for figure 5 (c)) corresponds to the bunch number that these
resonances belong to (these bumps correspond to the `waists' in the case
of the large number of oscillations (figures 4 (b) and (c)).

Note also that the dependences in figures 5 (a) and (c) are similar to
those obtained for grating eigenmodes in the case of zero variations of
the mean permittivity at the grating boundaries [1,2]. The main difference
from the case when $\Delta\epsilon_1 = \Delta\epsilon_2 =0$ [1,2] is
that the $x$ dependencies of the scattered wave amplitudes in figures
5 (a) and (c) are not close to zero at the grating boundaries. If this
were the case for a grating with constant mean permittivity, there
would have been substantial energy losses from the grating because
propagating scattered waves outside the grating carry the energy away
from the grating. As a result, the corresponding grating eigenmodes
would have quickly decayed owing to strong leakage. This is the reason
why only a limited number of grating eigenmodes can exist in the gratings
with constant mean permittivity, that is only those for which the
scattered wave amplitude is close to zero at the grating boundaries [1,2].

In the presence of the conventional guiding effect in the grating
(i.e. when $\Delta\epsilon_1 = \Delta\epsilon_2 <0$), the situation
is significantly different. In the scattering angle range
$\theta_{21\mathrm{int}}<\theta_{21} \le \theta_{21r}$, the scattered
waves outside the grating are non-propagating (evanescent) waves that
do not carry energy away from the grating. Therefore, the leakage of
the grating eigenmodes can occur only as a result of the 0 diffracted
order (incident wave) at the rear boundary, which has an amplitude that
is much smaller than the amplitudes of the incident and scattered waves
in the grating (recall that other diffracted orders in the considered
gratings have even smaller amplitudes). Therefore, the leakage is weak
in the whole mentioned range of angles $\theta_{21}$. As a result, a
number of new (additional), weakly leaking grating eigenmodes with
different field structures (presented, for example, in figures 4 (b)
and (c) and 5 (b) and (d)) can exist in a slab with a holographic grating.

As mentioned above, variations in the mean permittivity in the grating
region (in the slab) result in only limited variations in the angles
$\theta_{21rs}$ corresponding to strong GAS resonances in figures 2 (b)
and (c) and 3 (a)­(c). At the same time, these variations may have a
drastic effect on height and sharpness of the GAS resonances. This means
that changing the mean permittivity in the grating region
(slab) must have a substantial impact on the leakage of the corresponding
grating eigenmodes.

\begin{figure}[!tb]
\centerline{\includegraphics[width=0.7\columnwidth]{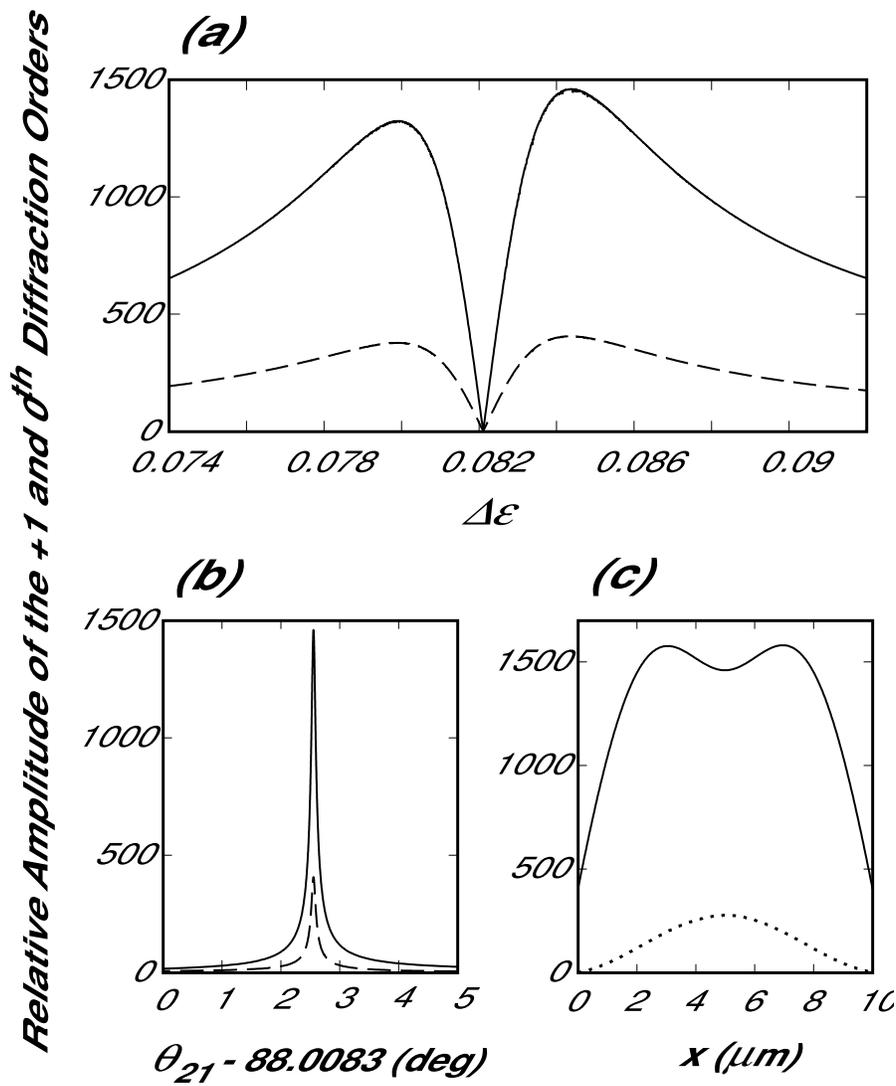}}
\caption{(a) The dependence of the maximal relative scattered wave
amplitude in the middle of the grating (---) for the GAS resonance at
$\theta_{21} \approx \theta_{21r}$ on the variation in the mean
permittivity at the grating boundaries:
$\Delta\epsilon = -\Delta\epsilon_1 = -\Delta\epsilon_2 >0$ (symmetric
structure). The broken curve gives the corresponding scattered wave
amplitudes at the front and rear grating boundaries (which are almost
the same). The other structural parameters are as follows:
$\epsilon_2 = 5$, $\epsilon_g = 0.03$, $\theta_{10} = 45^\circ$,
$L = 10 \mu$m and $\lambda(vacuum) = 1 \mu$m. (b) The dependences of
the scattered wave amplitudes in the middle of the grating (---) and at
the grating boundaries (-\,-\,-) on the scattering angle $\theta_{21}$
for the same structure as for figure 6 (a), but with
$\Delta\epsilon = 0.0844$, which is the optimized value of $\Delta\epsilon$
corresponding to the highest point of the solid curve in figure 6 (a).
(c) The $x$ dependences of the amplitudes of the scattered (---)
and incident (...) waves inside the grating at the optimized value of
$\Delta\epsilon = 0.0844$ and the resonant scattering angle
$\theta_{21} \approx \theta_{21r} \approx 88.008555^\circ$
(see figure 6 (b)).}
\end{figure}

It is not difficult to optimize the variations in the mean permittivity,
$\Delta\epsilon = -\Delta\epsilon_1 = -\Delta\epsilon_2 >0$, so that a
particular GAS resonance becomes as high as possible for a given grating
amplitude. To do this, we plot the dependence of the scattered wave
amplitude (e.g. in the middle of the grating) on the scattering angle
$\theta_{21}$ and record the maximal value of this amplitude, corresponding
to a particular resonance for a given $\Delta\epsilon$. Then, change
$\Delta\epsilon$ and repeat the above procedure. As a result, we obtain
the dependence of maximal values of the scattered wave amplitude,
corresponding to a particular resonance, on $\Delta\epsilon$. For example,
the solid curve in figure 6 (a) represents such a dependence for the
maximal scattered wave amplitude in the middle of the grating for the
first GAS resonance (i.e. at $\theta_{21} \approx \theta_{21r}$).
The grating parameters are $\epsilon_g = 0.03$, $L = 10 \mu$m,
$\theta_0 = 45^\circ$ and $\lambda(vacuum) = 1 \mu$m. The broken curve
in this figure gives the corresponding values
of the scattered wave amplitudes at the front and rear grating boundaries.
Note that, since $\theta_{21r}$ weakly depends on $\Delta\epsilon$,
its values are slightly different for different points on the curves
in figure 6 (a). Thus, it is clear that variations in the mean permittivity
in the grating region (compared with the regions outside the grating) may
lead to a very substantial increase of the height of the GAS resonance
(figure 6 (a)).

The predicted strong resonances at optimal values of $\Delta\epsilon$
($\Delta\epsilon \approx 0.0844$ in figure 6 (a)) obviously result in
the generation of very weakly leaking grating eigenmodes. Thus,
optimization of $\Delta\epsilon$ to the strongest GAS resonance
automatically means optimization of the corresponding eigenmode
to the minimal
losses due to leakage. According to the results of the rigorous analysis
of nonsteady-state EAS [16], the typical propagation distance for such an
eigenmode is expected to be around several tens of metres. However, the
exact analysis of this question will require more consistent application
of the methods developed in [16] to GAS in the considered structure
(which is beyond the scope of the current paper).

The dependences of
the relative scattered wave amplitudes on angle of scattering at the
optimized value of ($\Delta\epsilon \approx 0.0844$ (figure 6 (a)) are
presented in figure 6 (b) by the solid curve (in the middle of the grating)
and broken curve (at the grating boundaries). It can be seen that the
angular width of the optimized resonance is about $1.3\times 10^{-5}{}^\circ$.
It is quite clear that such a resonance can hardly be achieved in practice,
owing to the excessively long relaxation time. Therefore, its major
importance is in the demonstration of the existence of new, very weakly
leaking grating eigenmodes. In the end, these eigenmodes can be generated
by the same means as conventional slab modes (but not necessarily by means
of GAS).

The $x$ dependences of the amplitudes of the incident (dotted curve) and
scattered (solid curve) waves in the optimized grating eigenmode
(i.e. corresponding to the resonance in figure 6 (b):
$\Delta\epsilon \approx 0.0844$ and $\theta_{21} = 88.008555^\circ$)
are presented in figure 6 (c). It can again be seen that, unlike the
eigenmodes in gratings with zero variations in the mean permittivity
[1­3], the amplitudes of the scattered wave at the grating boundaries
are strongly non-zero. As has been indicated, this does not result in
any energy losses from the grating, since the scattered waves outside
the grating are evanescent. On the contrary, the amplitudes of the 0
diffracted order (incident wave) at the grating boundaries are negligible
compared with that in the middle of the grating (dotted curve in figure 6
(c)), which ensures only small leakage losses from the grating.

\section{Grating eigenmodes in planar waveguides}

As indicated in [4,6,11], a very similar pattern of scattering in the
geometry of EAS and GAS can also be obtained in the case of scattering
of guided slab modes of arbitrary polarization (TE or transverse magnetic
(TM) modes) in wide periodic groove arrays. In this case, the plane of
figure 1 is the surface of the slab with a periodic groove array within
a strip of width $L$. An incident guided mode of the slab (with the wave
vector $\mathbf{k}_{00}$) propagates at an angle $\theta_{10}$ (figure 1),
enters the array and is scattered into a scattered mode of the same slab,
so that the wave vector $\mathbf{k}_{21}$ of the scattered mode is almost
parallel to the array boundaries (the geometry of GAS) (figure 1).
The mean thickness of the waveguide inside the array (at $0<x<L$) is
assumed to be larger than outside it, that is, the mean thickness of
the slab experiences a step-like variation at the grating boundaries.
This results in a step-like increase in the effective mean permittivity
for the guided modes inside
the grating region (similar to what was considered in sections
3 and 4 for bulk waves).

The procedure of extending the results obtained in the above sections for
bulk TE electromagnetic waves in holographic gratings to the case of GAS
of guided slab modes in periodic groove arrays is presented and justified
in [4,11]. In particular, it follows from here that slanted periodic groove
arrays are capable of guiding electromagnetic modes, similar to the
eigenmodes of regular rib waveguides. At the same time, these grating
eigenmodes are strongly different in terms of field structure and existence
conditions from the conventional modes of a rib waveguide (see section 4).
The field structure of the grating eigenmodes of a rib with a periodic
groove array must be approximately the same as that described in section
4 (for further justifications see [4,11]). Similar to the considered
eigenmodes of a slab with a holographic grating (section 4), eigenmodes
of a rib waveguide with a periodic groove array must be weakly leaking
into guided modes in the regions outside the array. The sharper the
corresponding GAS resonance, the weaker is this leakage, and the longer
is the propagation distance for the eigenmodes.

However, it is important
to understand that the extension of the obtained results (sections 3 and 4)
to the case of guided modes in periodic groove arrays has only been justified
within the frames of the applicability of the approximate theory of GAS,
based on the two-wave approximation [4,11]. The actual applicability
condition for the approximate theory of GAS of bulk electromagnetic waves
was derived in [1] and verified in [2,8]:
\begin{equation}
\rho_\mathrm{gas} = \frac{2\lambda^2}{\Lambda^2} \left|
\frac{E_{00}}{\epsilon_g \mathrm{max}|S_{21}(x)|} \right | > 10,
\end{equation}
where $\lambda$ is the wavelength in vacuum, $S_{21}(x)$ is the amplitude
of the $+1$ order (scattered wave) inside the grating and $\epsilon_g$
is the amplitude of the holographic grating.

If condition (2) is satisfied
for GAS of bulk TE waves in a holographic grating, then we can find a
structure with GAS of guided modes in a slab with a periodic groove array
[4,11], for which all the dependences of the amplitudes of the incident
and scattered waves are very close to those for bulk TE waves (sections
3 and 4). Similarly, if we have a particular structure with GAS of guided
modes (without mode conversion), the same procedure [11] allows us to find
an equivalent structure with bulk TE electromagnetic waves in a holographic
grating with the same dependences of the incident and scattered wave
amplitudes.

Another applicability condition for the discussed extension to GAS of
guided modes is that the corresponding variations of the mean thickness
of the slab at the boundaries of the groove array should be much smaller
than the mean thickness itself [4,11]. If this is not the case, significant
energy losses from the structure due to generation of bulk waves at the
array boundaries may occur, resulting in deterioration of the predicted
resonances.

One might also think that owing to very large amplitudes of
the scattered wave at the grating boundaries (figures 4 (b), 5 (d) and
6 (a)­(c)), even small variations in the mean thickness of the slab may
result in the significant energy losses. However, this is not the case,
since GAS resonances occur at scattering angles
$\theta_{21\mathrm{int}}< \theta_{21} \le \theta_{21r}$. Within this range,
the $y$ component of the wave vector of the scattered
guided mode in the array is larger than the wave vector of the same
mode outside the array and, even more so, larger than the wave vector
of the bulk waves in the media surrounding the slab. Therefore, propagating
bulk waves will not be generated at the grating boundaries by means of the
scattered slab mode in the range of angles where strong GAS resonances
occur.

On the contrary, the incident slab mode can result in generation of bulk
waves at the front and rear grating boundaries. However, this wave has
much smaller amplitudes at these boundaries, than the amplitude of the
scattered wave in the grating. Thus, the resultant additional energy
losses must be negligible for small variations in the mean thickness
of the slab. Small thickness variations correspond to small variations
in the mean effective dielectric permittivity for the guided modes, which
is in agreement with the typical (small) values of $\Delta\epsilon$
considered for bulk waves in sections 3 and 4.

\section{Conclusions}

This paper has used the rigorous theory [2,8,11] for the analysis of GAS
of bulk electromagnetic waves in wide holographic gratings with step-like
variations in the mean dielectric permittivity at the grating boundaries.
A highly unusual pattern of strong multiple wave resonances has been
predicted in such gratings when the scattered wave propagates nearly
parallel to the grating boundaries (the geometry of GAS). These resonances
are shown to exist only in gratings with sufficiently large amplitudes
and/or widths, when the mean permittivity in the grating region is larger
than the permittivities outside the grating. In this case, the grating
region is actually a slab capable of guiding electromagnetic modes.

At the same time, it is clearly demonstrated that the predicted GAS
resonances are unrelated to generation of conventional guided modes
of the slab. It has been shown that these resonances are associated
with generation of a completely new type of modes in a slab with a
periodic grating. In particular, it has been shown that these grating
eigenmodes have a different field structure and are generated at angles
of scattering that are substantially different from those required for
generation of conventional slab modes. Moreover, the predicted grating
eigenmodes are generated in gratings, where effective generation of
conventional slab modes is not possible due to relatively large grating
amplitude and/or width.

It has been demonstrated that the grating eigenmodes of a slab with a
holographic grating are intrinsically associated with scattering and
cannot exist in the absence of the grating. Unlike conventional slab
modes, they are formed not only by the scattered wave (the $+1$ order),
but also by a superposition of all diffracted orders generated in the
grating. Grating eigenmodes have been shown to leak weakly into the
regions outside the grating. The main existence conditions for grating
eigenmodes have been discussed. However, the higher the considered GAS
resonance, the longer is the propagation distance for the corresponding
grating eigenmode.

Optimization of the variations in the mean permittivity at the grating
boundaries has been made to achieve the highest GAS resonance(s) (which
can reach thousands of amplitudes of the incident wave at the front
grating boundary). This procedure automatically means optimization of
the corresponding eigenmodes(s) to the minimal leakage and thus maximal
propagation distance along the grating.

Extension of the obtained results to the case of GAS of electromagnetic
modes guided by a slab with a slanted periodic groove array has been
carried out. In particular, the existence of a new type of eigenmodes
of a rib waveguide with a slanted periodic groove array has been predicted.

\section*{References}

\begin{enumerate}
\item  Gramotnev, D.K., Opt. Quant. Electron., 2001. 33: p. 253.
\item  Gramotnev, D. K., T. A. Nieminen, Optics Commun., 2003. 219: p. 33. 
\item  Pile, D. F. P., D. K. Gramotnev, Applied Physics B, 2003. 76: p. 65. 
\item  Gramotnev, D. K., S. J. Goodman, D. F. P. Pile, J. Mod. Opt., 2003 (to be published).
\item  Kishino, S., A. Noda, and K. Kohran, J. Phys. Soc. Japan, 1972. 33: p. 158.
\item  Bakhturin, M.P., L.A. Chernozatonskii, and D.K. Gramotnev, Applied Optics, 1995. 34: p. 2692.
\item  Gramotnev, D.K., J. Phys. D, 1997. 30(14). p. 2056-2062.
\item  Nieminen, T.A. and D.K. Gramotnev, Opt. Commun., 2001. 189: p. 175.
\item  Gramotnev, D.K. and D.F.P. Pile, Opt. Quant. Electron., 2000. 33: p. 1097.
\item Pile, D. F. P., D. K. Gramotnev, Opt. Quant Electron., 2003. 35: p. 237. 
\item Gramotnev, D.K., T.A. Nieminen, and T.A. Hopper, J. Mod. Opt., 2001.
\item Akhmediev, N.N. and A. Ankiewicz, Solitions: Nonlinear pulses and beams. 1997: Chapman and Hall.
\item Moharam, M.G., et al., J. Opt. Soc. Am., 1995. A12: p. 1077.
\item Moharam, M.G., et al., J. Opt. Soc. Am., 1995. A12: p. 1068.
\item Yariv, A. and P. Yeh, Optical waves in crystals. Propagation and control of laser radiation. 1984, John Wiley, New York. 
\item Nieminen, T. A. and D. K. Gramotnev, Optics Express, 2002, 10, p.2683. 
\end{enumerate}

\end{document}